\documentclass[nologo,11pt,a4paper]{ETHpaper}
\usepackage{graphicx, amsmath, amssymb,color,wasysym}
\usepackage{bm}
\usepackage[square,numbers,sort&compress]{natbib}
\usepackage{caption}
\usepackage{subfigure}
\usepackage{longtable}
\usepackage{booktabs}
\usepackage{mathtools}
\usepackage{cancel,cool}

\begin{document}

\newcommand{\mean}[1]{\left\langle #1 \right\rangle} 
\newcommand{\abs}[1]{\left| #1 \right|} 
\newcommand{\ul}[1]{\underline{#1}}
\renewcommand{\epsilon}{\varepsilon} 
\newcommand{\eps}{\varepsilon} 
\renewcommand*{\=}{{\kern0.1em=\kern0.1em}}
\renewcommand*{\-}{{\kern0.1em-\kern0.1em}} 
\newcommand*{\+}{{\kern0.1em+\kern0.1em}}

\newcommand{\RA}{\Rightarrow}
\newcommand{\bbox}[1]{\mbox{\boldmath $#1$}}

\title{Neighborhood approximations for non-linear voter models}

\titlealternative{Neighborhood approximations for non-linear voter models}

\author{Frank Schweitzer$^{\star}$,\footnote{Corresponding author: \url{fschweitzer@ethz.ch}}
Laxmidhar Behera$^{\dagger}$ 
}

\authoralternative{Frank Schweitzer, Laxmidhar Behera}

\address{$^{\star}$Chair of Systems Design, ETH Zurich, Weinbergstrasse 58, 8092 Zurich, Switzerland \\
$^{\dagger}$ Department of Electrical Engineering, Indian Institute
    of Technology, Kanpur 208 016, India
}

\reference{\emph{Entropy}, vol. \textbf{15} (2015) 7658-7679}

\www{\url{http://www.sg.ethz.ch}}

\makeframing
\maketitle

\begin{abstract}
Non-linear voter models assume that the opinion of an agent depends on the opinions of its neighbors in a non-linear manner.
This allows for voting rules different from majority voting. 
While the linear voter model is known to reach consensus, non-linear voter models can result in the coexistence of opposite opinions. 
Our aim is to derive approximations to correctly predict the time dependent dynamics, or at least the asymptotic outcome, of such local interactions. 
Emphasis is on a  probabilistic approach to decompose the opinion distribution in a second-order neighborhood into lower-order probability distributions. 
This is compared with an analytic pair approximation for the expected value of the global fraction of opinions and a mean-field approximation. 
Our reference case are averaged stochastic simulations of a one-dimensional cellular automaton. 
We find that the probabilistic second-order approach captures the dynamics of the reference case very well for different non-linearities, i.e for both majority and minority voting rules, which only partly holds for the first-order pair approximation and not at all for the mean-field approximation.
We further discuss the interesting phenomenon of a correlated coexistence, characterized by the formation of large domains of opinions that dominate for some time, but slowly change.

  \emph{Keywords:} opinion dynamics; voter model; pair approximation; higher-order probability distribution, cellular automata  
\end{abstract}

\section{Introduction}

The concept of \emph{entropy} plays a key role in describing the transition between disorder and order. 
Specifically,  the emergence of order in a random phase is indicated by a significant reduction of entropy. 
Calculating this reduction requires to know the probability of each possible system configuration that is compatible with the given system constraints - a challenging problem both methodologically and computationally. 
If we consider a system of $N$ elements each of which can be in one of two states -- for example up and down spins in a physical system, or agents with opposite opinions in a social system, or agents with the two strategies \emph{cooperate} or \emph{defect} in an economic system -- the number of possible configurations is $2^{N}$, which can be quite large. 
In fact, statistical physics was founded in the 19th century to provide an efficient solution based on the concept of statistical ensembles and state sums. 

\begin{figure}[htbp]
  \centering
  \includegraphics[width=0.5\textwidth]{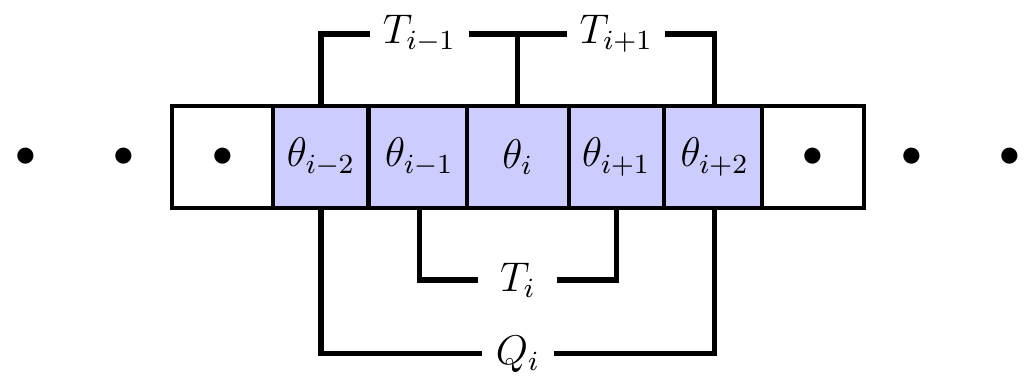}
  \caption{One-dimensional CA with triplet ($T_{i}$) and quintuplet ($Q_{i}$) definition of the neighborhood of agent $i$}
  \label{fig:1dca}
\end{figure}

In this paper, we address the problem by proposing a stochastic approach that allows to decompose such probabilities for systems characterized by  neighbor-neighbor interactions. 
Our candidate model to describe this interaction is the so-called \emph{voter model} which is discussed in more detail in Sect. \ref{sec:voter}. 
In this model, agents are in one of two discrete states, $\sigma\in\{0,1\}$, denoted as ``opinions''. 
They change their opinion  in response to the opinions in their neighborhood. 
In order to define such a neighborhood, we have chosen a one-dimensional cellular automaton (CA), in which consecutively numbered cells $i=1,...,N$ represent agents (see Figure \ref{fig:1dca}).   
We assume that the CA forms a ring to close the system.
Each agent $i$ then has two neighbors $i-1$, $i+1$ i.e. their opinions $\theta_{i}$ form a \emph{triplet} $T_{i}=\{\theta_{i-1},\theta_{i},\theta_{i+1}\}$. 
The second-order neighborhood that also takes the neighbors of $i-1$, $i+1$ into account, then results in a \emph{quintuplet} of opinions $Q_{i}=\{\theta_{i-2},\theta_{i-1},\theta_{i},\theta_{i+1},\theta_{i+2}\}$. 

Our major assumption of the voter model is that changes in the opinion of agent $i$ are only caused by the first-order neighborhood.
Specifically, the agent responds to the \emph{local frequency} of opinions in the triplet that also includes its own opinion. 
However, because the dynamics of agent $i$ depends on its neighbors, it is also coupled to their dynamics, i.e. to the second-order neighborhood of agent $i$, and so forth. This specifically denotes the problem that we are going to discuss in this paper.
In a stochastic approach, we are faced with a system of $N$ coupled dynamic equations for the probabilities $p(\theta_{i},t)$ to find any agent $i$ with opinion $\theta_{i}$ at time $t$. 
This coupling exists only through the two neighbors of each agent, i.e. changes in a far distant cell only propagate slowly through the CA by means of neighbor-neighbor interactions. 
So, precisely, how large should be the neighborhood taken into account for the dynamics of the CA? 
Or, how far reaching are  correlations in the changes of opinions? 

To answer this question, in this paper we propose three different analytic approximations for the dynamics which take different neighborhood sizes into account: zero-order -- no correlations between neighboring agents, first-order -- correlations between an agent and one of its nearest neighbors (so called pair approximation), second-order --  correlations between an agent and its second-nearest neighbors. 
We test the validity of these dynamic approximations by comparing them to stochastic simulations of the CA, averaged over a number of runs. 

Our emphasis is of course on the second-order approximation, which extends previous investigations. 
But we want to understand under which circumstances this approximation fares better than the simpler ones. 
Therefore, we have chosen different variants of the voter model, also known as \emph{nonlinear voter models}. 
The non-linearity is  with respect to the response to the local frequency of opinions. 
We further compare deterministic and stochastic dynamics, to show whether our analytic approximations can correctly predict the dynamics. 
Our variable of interest is the expected fraction of opinion 1, denoted as $\mean{x(t)}$, which is an aggregated variable for which we derive the dynamics based on a stochastic approach. 

Our paper combines, and extends, previous investigations in different directions. 
One line of research refers to the approximation of higher-order probability distributions by means of lower-order distributions \citep{brown1959note} which has been also applied to the voter model \citep{muehlenb-hoens-02}. 
This implies a loss of (microscopic) information which can be quantified by means of information-theoretic measures \citep{pfante2014comparison}. 
The question whether the coarse-grained dynamics is still Markovian can be answered by analyzing emergent macroscopic memory effects using information-theoretic measures \citep{gornerup2008method}. 
Such measures have also been applied to voter models \citep{banisch2014microscopic}.
In most cases, the Markov chain analysis becomes quite cumbersome and therefore is restricted to one-dimensional CA \cite{agapie2004markov}. 

Another line of research considers different forms of neighborhood approximations specifically for binary state-dynamics (see \cite{gleeson2013binary} for a good overview), which also has been applied to the voter model, already \cite{Schweitzer2009}. 
We note that our pair approximation approach follows \cite{Schweitzer2009}, but applies it here to a one-dimensional CA, which results in different expressions for $\mean{x(t)}$ and the correlations $c_{1|1}(t)$.

Compared to two-dimensional CA \cite{krause2012mean,stauffer-02acs} or even complex networks \cite{maxi2,suchecki2005}, one could find one-dimensional CA too simple. 
But this judgment is in fact not justified.
Already one-dimensional CA have proven to exhibit a really complex dynamics, with a chance to derive analytic expressions. 
Extensions of the simple voter model, for example the Sznajd model \cite{lb-fs-03} or the $q$-voter model \cite{przybyla2011exit}, could be thoroughly analyzed for one-dimensional CA. 

The emphasis of our investigations is on the validity of the analytic approximations for \emph{nonlinear} voter models. 
This non-linearity can be introduced in different ways. 
In Ref. \cite{stark2008slower,castellano2009nonlinear,xiong2013competition}, the authors discuss it on the level of individual agents that respond to neighboring influences in a \emph{heterogeneous} manner. Specifically, \cite{stark2008slower} assumes a heterogeneous inertia for agents to change their opinions, \cite{castellano2009nonlinear} assumes a heterogeneous neighborhood size to influence agents, while \cite{xiong2013competition} assumes a heterogeneous weight for the influence of agents. 

Compared to these approaches, we assume a \emph{homogeneous}, but nonlinear response of agents on the local frequency of opinions. 
Specifically, we consider the \emph{majority} rule, where the tendency of agents' to change their opinion \emph{increases} with the frequency of the opposite opinion. 
The \emph{minority} rule, on the other hand, assumes exactly the opposite, i.e. a \emph{decreasing} tendency. 
This can be simply varied by one parameter $\alpha$, which however is assumed to be the same for all agents. 

Eventually, we would like to point out that we refrain from interpreting our model in a social context. 
Although agents are called ``voters'' and their states are called ``opinions'', the simplicity of the underlying assumptions does not justify to sell the model as a reflection of a social system. 
We see it rather as a very generic setup to better understand the impact of local feedback on emerging systemic properties, such as \emph{consensus} (i.e. a ``ferromagnetic'' phase) or \emph{coexistence} (i.e. a ``paramagnetic'' phase). 
But we acknowledge that, despite this basic limitations, the voter model has been applied in various context, e.g. to model investors' behavior in financial markets \cite{e17052590}, emerging communication networks \cite{banisch2010opinion}, or invasion of species \cite{keitt01:_allee_effec_invas_pinnin_border}. 
A good overview of spin-type models in sociophysics is given in \cite{castellano2007rmp}. 

\section{Stochastic Approach }
\label{2}
\subsection{Defining the cellular automaton}
\label{2.1}

In this paper, we consider a one-dimensional cellular automaton (CA)
consisting of $N$ cells, each of which is identified by the index $i \in
N$ (see Figure \ref{fig:1dca}). Each cell shall be characterized by a discrete value
$\theta_{i}=\{0,1\}$, hence the total distribution of states is given by
the vector $
\mathbf{\Theta}=\{\theta_1,...,\theta_{i-2},\theta_{i-1},\theta_{i},
\theta_{i+1}, \theta_{i+2}, ...,\theta_N\}$. Assuming a torus space, each
cell $i$ has a clearly defined neigborhood of first and second nearest
neighbors, $\underline{\theta}_{i}'=\{\theta_{i-1},\theta_{i+1}\}$ and
$\underline{\theta}_{i}''=\{\theta_{i-2},\theta_{i+2}\}$. The probability
to find cell $i$ in state $\theta_{i}$ at time $t$ (where time shall be
measured in discrete steps) is $p(\theta_i,t)$. Consequently, the
conditional probabilities $p(\theta_i|\underline{\theta}_{i}^{'},t)$ and 
$p(\theta_i|\underline{\theta}_{i}^{'},\underline{\theta}_{i}^{''},t)$
describe the probability to find cell $i$ in state $\theta_{i}$ given
that it has the first and second nearest neighbors
$\underline{\theta}_{i}^{'}$, $\underline{\theta}_{i}^{''}$

Under Markov assumptions the Chapman-Kolmogorov equation holds for the
probability $p(\theta_{i},t)$ to find cell $i$ in state $\theta_{i}$ at time $t+1$:
\begin{equation}
  \label{eq:chap0}
  p(\theta_{i},t+1)= \sum_{\hat{\theta}_{i}\in \theta^{\star}} p \left[ (\theta_{i},t+1)   \gets (\hat{\theta}_{i},t)\right] \;p(\hat{\theta}_{i},t) 
\end{equation} 
The propagator $p [ (\theta_{i},t+1)   \gets (\hat{\theta}_{i},t)]$ denotes the transition probability to go from any given state $\hat{\theta}_{i}$ at time $t$ to the assumed state $\theta_{i}$ in the next time step, $t+1$. 
Here the summation is over all possible realizations $\theta^{\star}$ of
$\hat{\theta}_{i}$, i.e. the $2^{1}$ possible states $\{0,1\}$.  At this point,
we make our \emph{1st assumption}, namely that any change of $\theta_{i}$
depends on the \emph{nearest neighbors}, given by
$\underline{\theta}_{i}'=\{\theta_{i-1},\theta_{i+1}\}$. That means a \emph{triplet}
$T_{i}=\{\theta_{i-1},\theta_{i},\theta_{i+1}\}$ decides about the value of
$\theta_{i}$ in the next time step.  If we define the probability of a
triplet configuration as
\begin{equation}
  \label{eq:triplet}
  p(T_{i},t)=p(\theta_{i},\underline{\theta}_{i}',t)= 
p(\theta_{i-1},\theta_{i},\theta_{i+1},t)
\end{equation}
then the probability $p(\theta_i,t)$ results as the marginal
distribution of the triplet probability: 
\begin{equation}
  \label{eq:marginal}
  p(\theta_{i},t)=\sum_{\underline{\theta}^{\prime}_{i}\in \underline{\theta}'^{\star}} 
  p(\theta_{i},\underline{\theta}_{i}',t)
  = \sum_{T_{i}\in T^{\star}} p(\theta_{i}|T_{i},t)
\end{equation}
Here the summation is over all possible realizations
$\underline{\theta}'^{\star}$ of the nearest neighborhood
$\underline{\theta}_{i}'$, i.e. $2^{2}$ different possibilities.
$p(\theta_{i}|T_{i},t)$ denotes the conditional probability to find
$\theta_{i}$ as the focal cell given a triplet $T_{i}$, and the summation
is over all possible realizations $T^{\star}$ of $T_{i}$, i.e.  $2^{3}$
diffferent possibilities.

Based on the assumption that the nearest neighborhood matters, we can
rewrite Eqn. (\ref{eq:chap0}) as
\begin{equation}
  \label{eq:chap1}
p(\theta_{i},t+1)= \sum_{\hat{\theta}_{i}\in \theta^{\star}}
\sum_{\underline{\theta}^{\prime}_{i} \in \underline{\theta}'^{\star}}
p\left[(\theta_{i},t+1) 
\gets (\hat{\theta}_{i}|\underline{\theta}'_{i},t)\right] 
\;p(\hat{\theta}_{i}|\underline{\theta}'_{i},t) 
\end{equation} 
where $p(\theta_{i}|\underline{\theta}_{i}',t)$ denotes the
conditional probability to find the focal cell in state $\theta_{i}$
given the neighborhood $\underline{\theta}'_{i}$. Using Bayes' rule,
we can express this probability as:
\begin{equation}
  \label{eq:bayes0}
  p(\theta_{i}|\underline{\theta}_{i}',t)=
  \frac{p(\theta_{i},\underline{\theta}_{i}',t)}{p(\underline{\theta}_{i}',t)}
=\frac{p(T_{i},t)}{p(\underline{\theta}_{i}',t)}\;;\quad
\sum_{\underline{\theta}_{i}'\in \underline{\theta}'^{\star}}p(\underline{\theta}_{i}',t)=1
\end{equation}
With this, we can eventually rewrite the Chapman-Kolmogorov Eqn.
(\ref{eq:chap0}) for the single cell $i$ in terms of the triplet
probabilities, $p(T_{i},t)$:
\begin{equation}
  \label{eq:chap}
p(\theta_{i},t+1)= \sum_{T_{i}\in T^{\star}} p\left[(\theta_{i},t+1) 
\gets (T_{i},t)\right] \;p(T_{i},t) 
\end{equation}
The propagator $p[(\theta_{i},t+1) \gets (T_{i},t)]$ describes the
transition probabilities to go from any possible triplet $T_{i}$ to a
state $\theta_{i}$ during the next time step.

This equation leaves us with the further specification of the triplet
probability, $p(T_{i},t)$. While, according to our 1st assumption,
the occurence of $\theta_{i}$ is just determined by the nearest neighbors
$\underline{\theta}_{i}'=\{\theta_{i-1},\theta_{i+1}\}$, the occurrence
of either $\theta_{i-1}$ or $\theta_{i+1}$ also depends on their nearest
neighbors, i.e. the second nearest neighbors of $i$,
$\underline{\theta}_{i}''=\{\theta_{i-2},\theta_{i+2}\}$.  That means a
\emph{quintuplet}
$Q_{i}=\{\theta_{i-2},\theta_{i-1},\theta_{i},\theta_{i+1},\theta_{i+2}\}$ decides
about the value of the triplet $T_{i}=\{\theta_{i-1},\theta_{i},\theta_{i+1}\}$ in
the next time step. If we consider a quintuplet configuration $Q_{i}$,
then we can define the triplet probability as the marginal probability:
\begin{eqnarray}
  \label{eq:quad}
  p(Q_{i},t)&=&p(\theta_{i},\underline{\theta}_{i}',\underline{\theta}_{i}'',t)=
  p(\theta_{i-2},\theta_{i-1},\theta_{i},
  \theta_{i+1},\theta_{i+2},t) \\
  p(T_{i},t) &=& \sum_{\underline{\theta}'' \in \underline{\theta}''^{\star}}
  p(\theta_{i},\underline{\theta}_{i}',\underline{\theta}_{i}'',t) 
=\sum_{Q_{i}\in Q^{\star}} p(T_{i}|Q_{i},t)
\end{eqnarray}
Here the summation is over all possible realizations
$\underline{\theta}''^{\star}$ of the nearest neighborhood
$\underline{\theta}_{i}''$, i.e. $2^{2}$ different possibilities.
$p(T_{i}|Q_{i},t)$ denotes the conditional probability to find the
focal triplet $T_{i}$ given a quintuplet $Q_{i}$, and the summation is
over all possible realizations $Q^{\star}$ of $Q_{i}$, i.e.  $2^{5}$
diffferent possibilities.

For the dynamics for the probability to find a triplet in state $T_{i}$
at time $t+1$, we can write a Chapman-Kolmogorov equation quite similar
to Eqn. (\ref{eq:chap})
\begin{equation}
  \label{eq:chap2}
  p(T_{i},t+1)= \sum_{Q_{i}\in Q^{\star}} p\left[(T_{i},t+1) 
  \gets (Q_{i},t)\right] \;p(Q_{i},t) 
\end{equation}
The propagator $p[(T_{i},t+1) \gets (Q_{i},t)]$ describes the
transition probabilities to go from any possible quintuplet $Q_{i}$ to a
triplet $T_{i}$ during the next time step.

In order to specify the quintuplet probability $p(Q_{i},t)$, we may
consider a neighborhood of seven, etc. However, following the above
procedure repeatedly is neither convenient nor practicable since at the
end we have to consider $2^{N}$ possible configurations. Therefore, in
the next section, we present a more convenient method to determine the
quintuplet probability distribution.

\subsection{Quintuplet approximation}
\label{sec:app}

Instead of defining the quintuplet distribution in terms of higher-order
probability distributions, following the procedure of the previous
section, we now make our \emph{2nd assumption} by expressing these
probabilities in terms of \emph{lower-order distributions}, i.e. triplet
distributions, this way arriving at a closed form description of the
problem.

In the following we refer to the theory of approximating discrete
probability distributions already developed in the late 1950 \citep{brown1959note},
which was also applied to VM \citep{muehlenb-hoens-02}. The idea is to
approximate an $n$th order probability distribution, $p(x_{1},...,x_{n})$
by products containing the probabilities of given subsets,
e.g. $p(x_{i},x_{j},x_{k})$. It is known that such a product
approximation contains at most $n$ terms, since every new term has to
contain at least one variable $x_{i}$ not contained in previous
terms. For example, the factorization
$p(x_{1},...,x_{n})=\prod_{i}p(x_{i})$ already satisfies this condition,
although it may be not a good approximation, since it holds only for
ideal systems.

We note that the approximation procedure is not unique, i.e. there are
various ways of constructing a product approximation. For example, taking
$p(x_{1},x_{2},x_{3})$, this can be approximated by
$p(x_{1},x_{3})p(x_{2})$ as well as by $p(x_{1},x_{3})p(x_{2}|x_{1})$ or
by $p(x_{1},x_{3})p(x_{2}|x_{3})$.
Which one of the above approximations is the most suitable has to be
determined based on the information available about the triplet
distribution $p(x_{1},x_{2},x_{3})$. In some cases, there are
measurements of this distribution, or additional information about the
dependency between the $x_{i}$ as we will use below.

With reference to our quintuplet approximation using the triplets
defined, we propose the following procedure: We first note that there are
different triplets contained in the quintuplet configuration, for
example: $T_{i}=\{\theta_{i-1},\theta_{i},\theta_{i+1}\}$,
$T_{i-1}=\{\theta_{i-2},\theta_{i-1},\theta_{i}\}$,
$T_{i+1}=\{\theta_{i},\theta_{i+1},\theta_{i+2}\}$. These are valid
triplets because they consider the correct neighborhood relations,
whereas for example $\{\theta_{i-1},\theta_{i+1},\theta_{i+2}\}$ would be
inappropriate. Out of the valid triplets, we may choose for example the
first one, $\{\theta_{i-1},\theta_{i},\theta_{i+1}\}$ and complete the
product approximation as follows:
\begin{eqnarray}
  \label{eq:5app}
p(\theta_{i-2},\theta_{i-1},\theta_{i},
  \theta_{i+1},\theta_{i+2},t) = 
p(\theta_{i-1},\theta_{i},
  \theta_{i+1},t)\,p(\theta_{i-2}|\theta_{i-1},\theta_{i},t)
\,p(\theta_{i+2}|\theta_{i},\theta_{i+1},t)  
\end{eqnarray}
In order to express the conditional probabilities using the triplet
distribution, we apply Bayes' rule: 
\begin{equation}
  \label{eq:bayes}
  p(\theta_{i-2}|\theta_{i-1},\theta_{i},t)=\frac{p(\theta_{i-2},\theta_{i-1},\theta_{i},t)}{
    p(\theta_{i-1},\theta_{i},t)}
\end{equation}
which results in the final quintuplet approximation: 
\begin{eqnarray}
  \label{eq:5app-fin}
  p(\theta_{i-2},\theta_{i-1},\theta_{i},
  \theta_{i+1},\theta_{i+2},t) = 
p(\theta_{i-1},\theta_{i},
  \theta_{i+1},t)\,\frac{p(\theta_{i-2},\theta_{i-1},\theta_{i},t)}{
p(\theta_{i-1},\theta_{i},t)}\,
\frac{p(\theta_{i},\theta_{i+1},\theta_{i+2},t)}{
p(\theta_{i},\theta_{i+1},t)}
\end{eqnarray}
We note that this product approximation is the only possible one, given
that we have an ordered set of variables $\theta_{i}$ according to their
neighborhood relations. 

\subsection{Closed form dynamics}
\label{sec:dyn}

Using the approximation explained above, we have reduced the higher-order
description to the level of triplet distributions,
$p(\theta_{i-1},\theta_{i}, \theta_{i+1},t)$, etc., which results in a
closed form dynamics. We now have to specify the initial conditions for
the distributions. While we are able to make any suitable assumption
about the initial triplet distribution, we assume here that the
occupation of the different cells is initially statistically independent,
i.e. 
\begin{equation}
  \label{eq:equal}
  p(\theta_{i-1},\theta_{i}, \theta_{i+1},0)=
  p(\theta_{i-1},0)\;p(\theta_{i},0)\;p(\theta_{i+1},0)=p^{3}(\theta,0)
\end{equation}
For further investigations, we have set $p(\theta,0)=p(1-\theta,0)=1/2$. 

In order to solve the above set of equations, we have two different
possibilities: we can solve the dynamics (a) on the level of the
distribution $p(\theta_{i},t)$, Eqn. (\ref{eq:chap}), or (b) on the
level of the triplet distribution $p(T_{i},t)$,
Eqn. (\ref{eq:chap2}). Variant (a) requires more computational
effort, since we have to replace any $p(T_{i},t)$ as a marginal
distribution of triplets, whereas in variant (b), given the reduction of
the quintuplet distributions, we already have a closed form dynamics for
the triplet distributions, which is computationally more efficient.

Variant (b), however, requires to know the transition probabilities on
the level of the \emph{triplets}, whereas these are usually specified on
the level of single cell changes (see also the following section). Hence,
we have to determine the propagators $p[(T_{i},t+1) \gets
(Q_{i},t)]$ in terms of $p[(\theta_{i},t+1) \gets
(T_{i},t)]$. Since our 1st assumption specifies that the change of
every cell in the triplet only depends on its nearest neighbors, we are
able to factorize these transition probabilities: 
\begin{eqnarray}
  \label{eq:factor}
&  p\left[(T_{i},t+1) \gets (Q_{i},t)\right]=  p\left[(\theta_{i-1},t+1)
  \gets (T_{i-1},t)\right] \;\times   \nonumber \\   &
\; \times \; p\left[(\theta_{i},t+1) \gets (T_{i},t)\right] \; p\left[(\theta_{i+1},t+1) \gets
  (T_{i+1},t)\right]
\end{eqnarray}
This completes our closed form description.  To calculate the
time-dependent probability distribution of triplets, we have to insert
eqs. (\ref{eq:5app-fin}), (\ref{eq:factor}) into Eqn. (\ref{eq:chap2})
and from the result calculate $p(\theta_{i},t)$ as the marginal
distribution, Eqn. (\ref{eq:marginal}).  The only remaining task is now
to specify the transition probabilities for single cells,
$p[(\theta_{i},t+1) \gets
(T_{i},t)]$  at which point
the VM comes into play.

\section{Nonlinear Voter Dynamics}
\label{sec:voter}

\subsection{Transition rates}
\label{sec:trans}

The \emph{voter model}, in its basic form, is applied to a population of $N$ agents, where each agent is characterized by a discrete value, its \emph{``opinion''}, $\theta_{i}\in\{0,1\}$. 
The \emph{opinion dynamics} at the level of the agents works as follows: two agents $i$ and $j$ are \emph{randomly} chosen from the population, and agent $j$ adopts the opinion of agent $i$, i.e., $\theta_{j}(t+1)=\theta_{i}(t)$. After $N$ such update events, time is increased by 1 (random sequential update). 

The rather simplified rule  limits the applications of the voter model to any real voting process or opinion dynamics.
In a well mixed population, the probability for agent $j$ to adopt a given opinion $\sigma\in\{0,1\}$ is simply proportional to the global frequency $x_{\sigma}(t)$ of agents with the respective opinion:
  \begin{equation}
    \label{eq:1}
    x_{\sigma}(t)=\frac{N_{\sigma}(t)}{N}\;; \quad x(t)\equiv x_{1}(t)=1-x_{0}(t)\;; \quad N=\sum_{\sigma}N_{\sigma}(t)=N_{0}(t)+N_{1}(t)=\mathrm{const.}
  \end{equation}
This dynamics is known to always converge to complete \emph{consensus}, $x\to 1$ or $x\to 0$. 
The only interesting question is then how long it will take to reach this state dependent on the system size $N$ and the initial condition $x(0)$. 

This picture becomes more complex if instead of a well-mixed population a defined neighborhood for each agent is assumed. 
This can be a social network where agents have \emph{links} to other agents, or simply a lattice where the neighborhood is given by the geometry that defines the nearest and second nearest neighbors of an agent. 
Then, instead of the global frequency, the probability of an agent to adopt a given opinion depends on the \emph{local frequency}, $f_{i}^{\sigma}$, of this opinion in the neighborhood of that agent:
\begin{equation}
  \label{sum}
f_{i}^{\sigma} = \frac{1}{3}\left[\delta_{\sigma,\theta_{i-1}}+ \delta_{\sigma,\theta_{i}} + \delta_{\sigma,\theta_{i+1}}\right]
\;;\quad
f_{i}^{(1\-\sigma)} = 1 - f_{i}^{\sigma}
\end{equation}
Here $\delta_{x,y}$ means the Kronecker delta, which is 1 only for $x=y$
and zero otherwise. 
In this paper, we restrict ourselves to the one-dimensional regular lattice, where each agent is represented by a cell $i$ that has precisely two neighbors. 
Different from the basic voter rule, in the calculation of the local frequency we have taken the opinion $\theta_{i}$ of the focal cell $i$ into account.
This adds some inertia to the dynamics as agent $i$ counts towards the local minority/majority. 
It also avoids stalemate situations where the two neighbors have different opinions.

The linear voter model would assume that the transition rate
\begin{equation}
w(1\-\theta_{i}|\theta_{i}=\sigma,f_{i}^{\sigma})=f_{i}^{(1\-\sigma)}
\label{eq:2}
\end{equation}
of a focal cell $i$ to change its opinion from $\theta_{i}=\sigma$ to $1\-\theta_{i}=1\-\sigma$ is directly proportional to the local frequency $f_{i}^{(1\-\sigma)}$, i.e. the \emph{opposite} opinion in the neighborhood. 
The expression of Eq. (\ref{eq:2}) is a specification of the more general transition probability $p[(\theta_{i},t+1) \gets
(T_{i},t)]$ used in Eq. (\ref{eq:factor}). 
It takes into account that for frequency dependent processes the change of $\theta_{i}$ depends not directly on the local distribution
$T_{i}$, but only on the \emph{local frequency} $f_{i}^{(1\-\sigma)}$.

We are more interested in the \emph{non-linear} case, where the local frequency still plays a role, however the response to the opposite opinion can be different. 
The linear voter model can then be generalized into a \emph{majority} rule where the transition toward the opposite opinion \emph{monotonously increases} with the respective local frequency.
Different from this, the \emph{minority} rule would assume that the transition toward the opposite opinion \emph{monotonously decreases} with the respective local frequency, i.e. the minority opinion is favored. 
Eventually, we could also have \emph{mixed rules} with a \emph{non-monotonous} frequency dependence, which become possible only if the local frequency depends at least on the opinion of three agents, as it is the case in our model.

Instead of an analytical expression, we specify  the transition rate $w(1\-\theta_{i}|\theta_{i}=\sigma,f_{i}^{\sigma})$ by using free parameters $\alpha_{0}$, $\alpha_{1}$, $\alpha_{2}$, $\alpha_{3}$ which cover all of the above cases in a general manner. 
With these, the transition rates shall be defined as follows:
\begin{equation}
\begin{tabular}{ccc}
        $\quad f_{i}^{\sigma} \quad $ & $\quad f_{i}^{(1\-\sigma)} \quad $ &
        $w(1\-\theta_{i}|\theta_{i}=\sigma,f_{i}^{\sigma})$ \\ \hline
         1& 0 & $\alpha_{0}$\\
         2/3&1/3&$\alpha_1$\\
         1/3&2/3&$\alpha_2=1-\alpha_{1}$\\
         0 & 1 & $ \alpha_{3} = 1-\alpha_{0}$\\
        \end{tabular}
        \label{trans2}
\end{equation}
The general case of four independent transition rates in Eqn. (\ref{trans2}) can be reduced to two transition rates $\alpha_{0}$, $\alpha_{1}$ by assuming a symmetry between the two states $0$ and 1, i.e. $\alpha_{1}+\alpha_{2}=1$, $\alpha_{0}+\alpha_{3}=1$. 
The condition 
\begin{equation}
  \label{eps-alpha}
\alpha_{0} \leq \alpha_1 \leq (1-\alpha_{1}) \leq (1-\alpha_{0})\;;\quad \alpha_{1} \leq 1/2
\end{equation}
then denotes the above mentioned \emph{majority rule}, because the transition rate 
\emph{increases} with an increasing fraction of the
\emph{opposite} opinion $(1\-\sigma)$ in the neighborhood. This process
is also known as \emph{positive freqency dependent invasion}. 
In ecology, it means that individuals of abundant species have a better chance to survive.

Opposite to that, for the so-called \emph{negative freqency dependent invasion}
process the probability that cell $i$ shall switch to  the opposite opinion $(1\-\sigma)$ \emph{decreases} with the fraction number of
individuals of subpopulation $\sigma$ in the neighborhood. This implies
\begin{equation}
  \label{eps-alpha2}
(1-\alpha_{0}) > (1-\alpha_{1}) > \alpha_{1} > \alpha_{0}\;;\quad \alpha_{1}> 1/2 
\end{equation}
In an ecological context negative frequency dependent invasion means that
individuals of a rare subpopulation have a better chance to survive. 

The rate $\alpha_{0}$ in Eqn. (\ref{trans2}) applies for the case where cell $i$ is only surrounded by opinions of the same kind. In
a deterministic model, there would be no force to change the current 
state. In a stochastic CA however all possible processes should have a
certain non-zero probability to occur, therefore a rather small value
$\alpha_{0}\equiv\varepsilon= 10^{-4}$ is used to avoid absorbing states in the dynamics. 
We will then only vary the remaining rate $\alpha\equiv \alpha_{1}$, which is the only free parameter in the model.
Table summarizes the different transition rates given the possible local configurations $T_{i}$. 
It should be read as follows: given that the local configuration is e.g. $T_{i}(t)=\{001\}$ at time $t$, the probability to find $T_{i}(t+1)=\{011\}$ is $\alpha$.
If $\alpha\leq 1/2$ this transition rate increases with $f_{i}^{(1)}$, hence it denotes a \emph{majority voting} rule, otherwise it denotes a \emph{minority voting} rule. 
\begin{table}[htbp]
  \centering
  \begin{tabular}[c]{|c|c|c|c|c|}
\hline
    $\theta_{i}(t)=1$ & $\quad f_{i}^{(1)}\quad $ & $\quad f_{i}^{(0)}\quad$ & $w(0|1,f_{i}^{(1)})$ & $\quad t+1 \quad$ \\ \hline
010 &  1/3 & 2/3 & $1-\alpha$ & 000 \\
011 & 2/3 & 1/3 & $\alpha$ & 001 \\
110 & 2/3 & 1/3 & $\alpha$ & 100 \\
111 & 1 & 0 & $\varepsilon$ & 101 \\ \hline \hline
   $\theta_{i}(t)=0$ & $f_{i}^{(1)}$ & $f_{i}^{(0)}$ & $w(1|0,f_{i}^{(1)})$ & $t+1$ \\ \hline
000 &  0 & 1 & $\varepsilon$ & 010 \\
001 & 1/3 & 2/3 & $\alpha$ & 011 \\
100 & 1/3 & 2/3 & $\alpha$ & 110 \\
101 & 2/3 & 1/3 & $1-\alpha$ & 111 \\ \hline
  \end{tabular}
  \caption{Transition probabilities for $\theta_{i}$ given different local configurations. $\alpha\leq 1/2$ denotes \emph{majority voting} rules, $\alpha> 1/2$ denotes \emph{minority voting} rules.}
  \label{tab:trans}
\end{table}

Now that the  transition rates for the nonlinear voter model are specified, we have different ways to proceed:
(1) We can run stochastic computer simulations of the one-dimensional CA, to get some intuition about the dynamics and the role of the parameter $\alpha$. This will be described in the following Sect. \ref{2.3}. (2) We can also derive an approximate macroscopic dynamics for the global frequency $x(t)$, which will be done in Sections \ref{3.1}, \ref{4.1} using two different approximations. (3) Eventually, we can use the closed-form description already derived in Sect.~\ref{sec:dyn}, to numerically calculate $x(t)$. The stochastic simulations will denote our \emph{reference case}, used to compare the two different approximate macroscopic dynamics and the numerical calculations.

\subsection{Computer simulations of the CA model}
\label{2.3}

For a first insight into the dynamics, we have conducted stochastic
simulations of the CA described above. In this section, we will only
refer to particular runs, to show some snapshots of the dynamics,
while in Sect. \ref{sec:num} also averaged simulation results of the global variables
are discussed. 

For our simulations, we have used a one-dimensional CA of $N=640$ cells with periodic boundary conditions,
where all cells are simultaneously updated. The discrete time scale is defined by
\emph{generations}. We have checked that the main results
do not change if $N$ is increased to 6400. 
The initial
configuration of the CA refers to a homogeneous distribution (within
discrete limits) of both opinions, i.e. initially each cell is randomly
assigned one of the possible states, $\{0,1\}$, with a probability that is equal to the initial global
frequency $x(t=0)$. 
At each time step, the transition rates are
calculated for each cell according to Eqn. (\ref{trans2}) and the values are
compared with a random number \texttt{rnd} drawn from the intervall
$(0,1)$. If \texttt{rnd} is less than the calculated transition rate, the
respective transition process is carried out, otherwise the cell remains
in its current state. Since each transition only depends on the current
local configuration, memory effects are not considered here.

It follows from the above description that the case $\alpha_{0}=\varepsilon=0$ and
$\alpha=0$ refers to a \emph{deterministic}
positive majority voting rule, simply because the state of cell $i$
\emph{never} changes unless the two nearest neighbor
cells have adopted the same opinion. But then, it will \emph{always} change, such that all three cells have the same opinion. 
Similarly, a \emph{deterministic}
minority voting rule is described by $\varepsilon=0$,
and $\alpha=1$.

\begin{figure}[htbp]
  \begin{center}
\includegraphics[width=7.0cm]{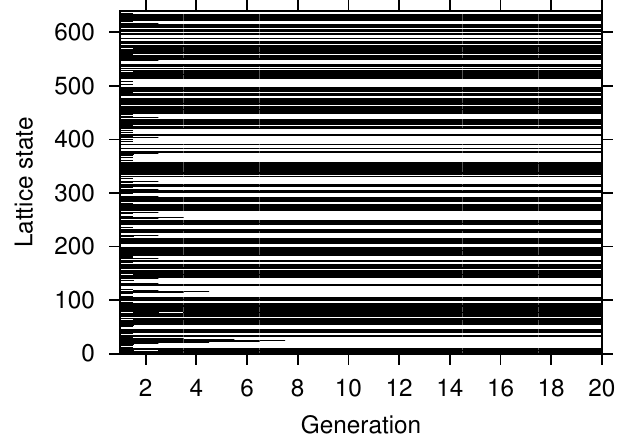}\qquad
\includegraphics[width=7.0cm]{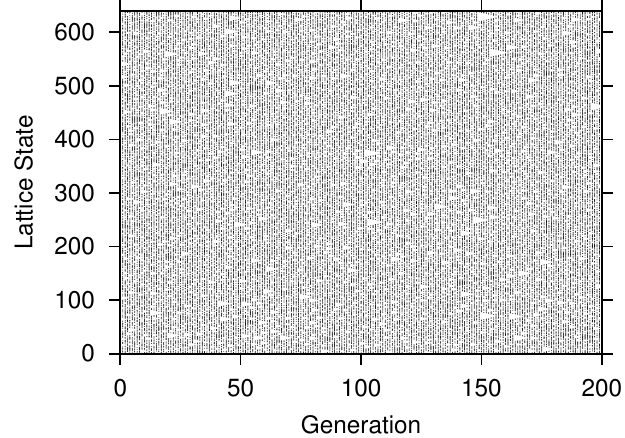}

\bigskip
\includegraphics[width=7.0cm]{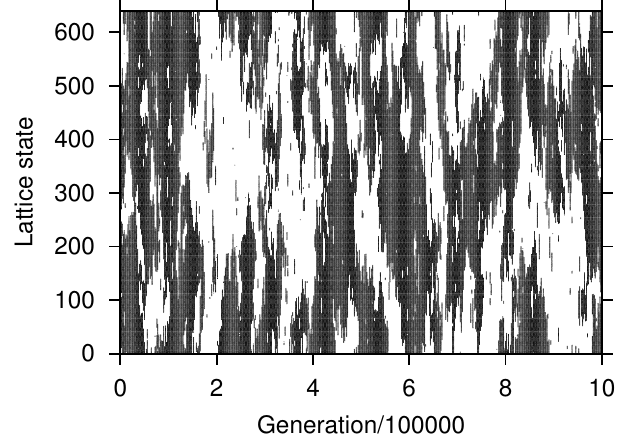}\qquad
\includegraphics[width=7.0cm]{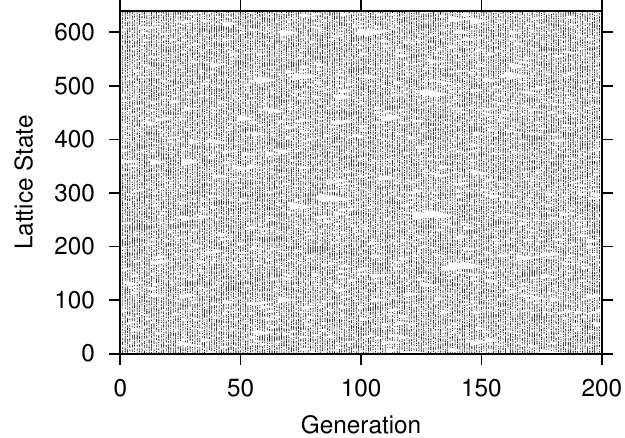}
  \end{center}
    \caption{Evolution of the CA for different rules: (left column) majority voting, (right column) minority voting, (top row) deterministic dynamics, (bottom row) stochastic dynamics. Parameters: (top left) $\varepsilon=0.0$, $\alpha=0.0$, (top right) $\varepsilon=0.0$, $\alpha=1.0$, (bottom left) 
 $\varepsilon=10^{-4}$, $\alpha=0.2$, (bottom right) $\varepsilon=10^{-4}$, $\alpha=0.8$. $N=640$, initial condition: $x(t=0)=0.5$, (black) indicates $\sigma=0$, (white) indicates $\sigma=1$. 
}
    \label{fig:ca1dp}
\end{figure}

The results of the computer simulations are illustrated in  Figure \ref{fig:ca1dp}. 
We first note the remarkable difference in the outcome between majority and minority voting.
In the latter, cells change their state frequently, however no pattern evolves because the cell always adopts an opinion different from the two neighbors, if these have the same opinion. 
Thus, we see a rather random and unstructured pattern, in which both opinions \emph{coexist} with a long-term fraction $x=0.5$. 
This overall behavior is also not changed in case of a stochastic dynamics. 

In contrast, for the majority voting rule we observe the formation of  local clusters of the same opinion. 
In the deterministic case, the resulting pattern becomes stationary very fast, such that the clusters remain rather small. 
In the stochastic case, however, we observe the interesting phenomenon that very large clusters of cells with the same opinion emerge. 
These domains do not grow up to system size because of the small perturbation $\varepsilon=10^{-4}$, that enables cells to adopt a different opinion even in a homogeneous neighborhood. 
The non-linearity ($\alpha=0.2$ is different from the linear case $\alpha=1/3$) then amplifies such random deviations.
This phenomenon has been discussed, for the case of a two-dimensional CA, in \citep{Schweitzer2009}. 
As a consequence, we observe a \emph{correlated coexistence} of both opinions characterized by the facts that (i) each opinion at times is the majority and (ii) forms larger clusters, and (iii) the dynamics is always in non-equilibrium. 
This is clearly visible in the long-term behavior, shown in  Figure \ref{fig:ca1dp} (bottom left). 

In Sect. \ref{sec:num}, we further investigate these different regimes numerically, after we have derived some appropriate approximations of the macroscopic dynamics.

\subsection{Derivation of the macroscopic dynamics}
\label{3.1}

A formal description for the dynamics of the CA starts with the probability $p(\theta_{i},t)$ used in Eqn. (\ref{eq:chap0}). 
By means of
\begin{equation}
  \label{eq:3}
  \mean{x_{\sigma}(t)}=\frac{1}{N}\sum_{i=1}^{N} p(\theta_{i}=\sigma,t)
\end{equation}
we can obtain the key variable of the macroscopic dynamics $\mean{x_{\sigma}(t)}$, which is the expected global frequency of each opinion in the population. 
Note that, different from Eq. (\ref{eq:1}) where $x_{\sigma}(t)$ is used, $\mean{x_{\sigma}(t)}$ describes the ensemble average over very many simulations. 
This is of some importance when interpreting the results. 
For the \emph{linear voter model}, it is known that the system converges always to consensus, i.e. $x\to 0$ or $x\to 1$. 
If we use the initial condition $x(0)=0.5$, then in 50\% of the cases we observe  $x\to 0$, and in 50\% $x\to 1$. 
However, when averaging over all of these outcomes, we find $\mean{x}=0.5$. 

Assuming a master equation for the dynamics of  $p(\theta_{i},t)$ with the transition rates specified in Eqn.~(\ref{trans2}), we can derive a rate equation for $\mean{x_{\sigma}}$ as discussed in detail in \citep{Schweitzer2009}: 
\begin{equation}
\label{eq:master_macrof}
\begin{aligned}
\frac{d}{dt} \mean{x_{\sigma}(t)} =  \sum_{\underline{\sigma}^{\prime}} \Big[
w(\sigma|(1\-\sigma),\underline{\sigma}^{\prime})\,
\mean{x_{(1\-\sigma),\underline{\sigma}^{\prime}}(t)}  - w(1\-\sigma|\sigma,\underline{\sigma}^{\prime})\,
\mean{x_{\sigma,\underline{\sigma}^{\prime}}(t)}\Big ]
\end{aligned}
\end{equation}
Here, the summation is over all possible $2^{n}$ opinion patterns $\underline{\sigma}^{\prime}$ for the neighborhood of cell $i$. 
These are binary strings $\sigma_{i-1}\sigma_{i+1}$ that indicate the particular values of the nearest neighbors $i-1$, $i+1$. 
With $n=2$ neighbors, these would be $00$, $01$, ${10}$, ${11}$. 
Together with the possible state for cell $i$, i.e. $\sigma=0$ or $\sigma=1$, the respective local frequencies $f_{i}^{\sigma}$ can be derived according to Eqn. (\ref{sum}), such that all transition rates in Eq. (\ref{eq:master_macrof}) are specified. 
The solution of Eqn. (\ref{eq:master_macrof}) would however require the computation of the averaged
global frequencies $\mean{x_{1,\underline{\sigma}^{\prime}}}$ and
$\mean{x_{0,\underline{\sigma}^{\prime}}}$ for all possible opinion patterns
$\underline{\sigma}^{\prime}$ over time, which would be a tremendous effort. 
Therefore, in \citep{Schweitzer2009} two analytic approximations have been discussed to solve this problem.
Here, we only summarize the results.  

In the first approximation, the so-called mean-field limit, the state of each cell does not depend on the
opinion of its neighbors, but is only influenced via a
mean field. In this case the opinion distribution factorizes:
\begin{equation}
  \label{factorize}
\mean{x_{\sigma,\underline{\sigma}^{\prime}}}= \mean{x_{\sigma}}\, \prod_{j=1}^{n}
\mean{x_{\sigma_j}} 
\end{equation}
and we find with $\mean{x}\equiv\mean{x_{1}}=1-\mean{x_{0}}$ for the macroscopic dynamics, Eqn. (\ref{eq:master_macrof})
in the mean-field limit:
\begin{equation}
\label{eq:master_macro_mean}
\begin{aligned}
\frac{d}{dt}\mean{x(t)}= \sum_{\underline{\sigma}^{\prime}}\Big[
w(1|0,\underline{\sigma}^{\prime})\, (1-\mean{x}) \prod_{j=1}^{n}
\mean{x_{\sigma_j}}  - 
w(0|1,\underline{\sigma}^{\prime}) \,\mean{x} \prod_{j=1}^{n}
\mean{x_{\sigma_j}}\Big]
\end{aligned}
\end{equation}
For the calculation of the $\mean{x_{\sigma_{j}}}$ we have to look at
each possible opinion pattern $\underline{\sigma}^{\prime}$. The
mean-field approach assumes that the occurence of each 1 or 0 in the
pattern can be described by the global frequencies $x$ and $(1\-x)$,
respectively  (for simplicity, the abbreviation $x\equiv\mean{x}$ will be
used in the following). Taking  the example $100$, i.e. $\sigma_{i}=0$ for cell $i$ and $\sigma_{i-1}=1$, $\sigma_{i+1}=0$ for its neighbors, would result in
$(1-\mean{x})\mean{x_{\sigma_{i-1}}}\mean{x_{\sigma_{i+1}}}= x(1\-x)^{2}$. 
Inserting further the
transition rates, Eqn. (\ref{trans2}), we find this way the equation for the
mean-field dynamics as: 
\begin{equation}
\label{eq:mean_field}
\begin{aligned}
  \frac{dx}{dt} = \varepsilon\left[(1-x)^3-x^3\right]+(1-3\alpha)\, x(1-x)\, (2x-1)
\end{aligned}
\end{equation}
The fixed points of the mean-field dynamics can be calculated from
Eqn. (\ref{eq:mean_field}) by means of $\dot{x}=0$. In the limit $\varepsilon=0$,
we find:
\begin{equation}
  \label{zero}
  \begin{aligned}
   x^{(1)} &= 0\;;\quad x^{(2)}= 1\;;\quad x^{(3)}=0.5
  \end{aligned}
\end{equation}
The three stationary solutions denote either consensus toward one of the opinions, or coexistence of both opinions with  an equal share. 
In order to verify the stability of the solutions, we have further
investigated the Jacobian $d/dx$ of Eqn. (\ref{eq:mean_field}). The results can
be concluded as follows: Below a critical reinforcement, $0<\alpha<1/3$,
$x=0$ and $x=1$ are stable attractors and $x=0.5$ is the separatrix. 
Above the critical reinforcement,  $1/3<\alpha<1$, $x=0.5$ becomes the stable attractor, while $x=0$ and $x=1$ are unstable. 
We will come back on this after discussing the second approximation.

\subsection{Pair Approximation}
\label{4.1}

The  mean-field approximation assumes that the local occurrence of opinions is determined by the \emph{global} frequencies rather than by local interactions. 
A better approximation should take local correlations between neighboring cells into account. 
Our second approximation, the so-called \emph{pair approximation}, is based
on the assumption that the state of each cell $i$ is only
correlated to the states of each of its nearest neighbors, separately. I.e., the two neighbors are only correlated through the focal cell and not to each other. 
Therefore, the neighborhood $\sigma_{i-1}\sigma_{i}\sigma_{i+1}$ is decomposed  in \emph{pairs} of
nearest neigbor cells, $\sigma_{i}\sigma_{i+1}$, $\sigma_{i-1}\sigma_{i}$ which are called
\emph{doublets}. The expected value of the  global frequency of
doublets is denoted as 
$\mean{x_{\sigma,\sigma^{\prime}}}$ where $\sigma$ refers to the focal cell and $\sigma^{\prime}$ to one of its neighbors.  

We can then approximate the global frequency of a specific
opinion pattern $\mean{x_{\sigma,\underline{\sigma}^{\prime}}}$ in Eqn. (\ref{eq:master_macrof}) as: 
\begin{equation}
\label{eq:pairapprox}
\mean{x_{\sigma,\underline{\sigma}^{\prime}}} =\mean{x_{\sigma}} \prod_{j=1}^{n}
  c_{{\sigma_j}|\sigma} 
\end{equation}
where the  $c_{\sigma|\sigma^{\prime}}$ are the correlations:
\begin{equation}
\label{eq:c-prob}
c_{\sigma|\sigma^{\prime}}:=\frac{\mean{x_{\sigma,\sigma^{\prime}}}}{\mean{x_{\sigma^{\prime}}}}
\end{equation}
that depend on the doublet frequency $\mean{x_{\sigma,\sigma^{\prime}}}$ and the global frequency of opinions $\mean{x_{\sigma^{\prime}}}$ 
neglecting higher order correlations.  $c_{\sigma|\sigma^{\prime}}$
can be interpreted  as the conditional probability that a
randomly chosen nearest neighbor of a cell in state $\sigma^{\prime}$ is
in state $\sigma$. With the 
relations:
\begin{equation}
\label{eq:cond_den}
\mean{x_{\sigma^{\prime}}}c_{\sigma|\sigma^{\prime}}=\mean{x_{\sigma}}
 c_{\sigma^{\prime}|\sigma} \;;
\sum_{\sigma^{\prime} \in \{0,1\}} c_{\sigma^{\prime}|\sigma}=1 
\end{equation}
and using again 
$\mean{x}\equiv\mean{x_{1}}$, these correlations can be expressed in
terms of only $c_{1|1}$ and $\mean{x}$ as follows:
\begin{equation}
\label{eq:cond_prob}
\begin{aligned}
c_{0|1}= 1-c_{1|1}\;;\quad
c_{1|0}= \frac{\mean{x}\, (1-c_{1|1})}{1-\mean{x}}\;; \quad
c_{0|0}= \frac{1-2\mean{x} +\mean{x}c_{1|1}}{1-\mean{x}}
\end{aligned}
\end{equation}
We then find for the macroscopic dynamics, Eqn. (\ref{eq:master_macrof}), in pair
approximation:
\begin{equation}
\label{macro-pair}
\begin{aligned}
\frac{d}{dt}\mean{x(t)}= \sum_{\underline{\sigma}^{\prime}}\Big[
w(1|0,\underline{\sigma}^{\prime})\, (1-\mean{x}) \prod_{j=1}^{n-1}
c_{\sigma_{j}|\sigma} - 
w(0|1,\underline{\sigma}^{\prime}) \,\mean{x} \prod_{j=1}^{n-1}
c_{\sigma_{j}|(1-\sigma)} \Big]
\end{aligned}
\end{equation}
With respect to Eqn. (\ref{eq:cond_prob}), Eqn. (\ref{macro-pair}) now depends on two
variables, $\mean{x}$ and $c_{1|1}$.
In order to derive a closed description, we need an additional equation
for $\dot{c}_{1|1}$, that can be obtained from Eqn. (\ref{eq:c-prob}):
\begin{equation}
\label{eq:local1}
\frac{dc_{1|1}}{dt} =  - \frac{c_{1|1}}{\mean{ x}}
\frac{d}{dt}\mean{ x}+\frac{1}{\mean{ x}}
\frac{d}{dt}\mean{ x_{1,1} }
\end{equation}
Eqn. (\ref{eq:local1}) requires additionally the time derivative of the global doublet frequency $\mean{x_{1,1}}$. 
We note that the three coupled equations for $\mean{x}$, $c_{1|1}$ and $\mean{x_{1|1}}$ can be easily solved numerically. 
In the Appendix, we have derived explicit expressions for these equations for the one-dimensional CA discussed here, using the transition rates of Eqn. (\ref{trans2}). 

\section{Numerical comparison of the approximations}
\label{sec:num}

\subsection{Computational procedure}
\label{sec:comp-proc}

In this section, we aim at a comparison of the different approaches developed so far. 
We recall that our \emph{reference case} is the stochastic simulation of the CA, where some sample runs were already presented in Sect. \ref{2.3}. 
To allow a comparison with the other approaches that predict the expected behavior, we will average the CA simulations over 50 independent runs. 
The initial condition is given by the global fraction of opinion 1, $x(t=0)$, which is realized as a uniform random distribution of the opinions as described already in Sect. \ref{2.3}. Thus, the initial value for the pair correlation is $c_{1|1}(t=0)=x(t=0)$. 
To decide when the CA simulations have reached stationarity, we verify that changes of  $\mean{x(t)}$ are less than $1/\sqrt{N}$. 

The dynamics of the reference case (a) shall be compared to the two other dynamic approximations, (b) quintuplet approximation (see Sect. \ref{sec:dyn}) and (c) pair approximation (see Sect. \ref{4.1} and Appendix). 
To facilitate the understanding, we summarize here the computational procedure. 

For the \emph{quintuplet approximation} our starting point is Eq. (\ref{eq:chap2}) which contains two terms that need a further approximation.
The quintuplet probability $p(Q_{i},t)$ can be decomposed in terms of triplet probabilities by means of Eq. (\ref{eq:5app-fin}).
The propagator $p[(T_{i},t+1)\gets (Q_{i},t)]$ can be decomposed in terms of propagators $p[(\theta_{i},t+1)\gets (T_{i},t)]$ by means of Eq. (\ref{eq:factor}). 
The latter are given as the transition rates of the VM in Table \ref{tab:trans}. 

Let us illustrate the computation by one example. 
A quintuplet has five states and hence $2^5=32$ configurations denoted as $Q^{\star}$ in  Eq. (\ref{eq:chap2}).
Each of these possible configurations will contribute to the triplet distribution according to Eq. (\ref{eq:chap2}).
Let us for the quintuplet take the sample configuration $\{01101\}$ and for the triplet $\{000\}$.
The right-hand side of Eq. (\ref{eq:chap2}) then reads: 
\begin{align}
  \label{eq:4}
  p[(0,t+1)\gets (011,t)]\  p[(0,t+1)\gets (110,t)]\  p[(0,t+1)\gets (101,t)] \times \nonumber \\
 \times \frac{p(011,t)\ p(1 1 0,t) \ p(1 0 1,t)}{p(1 1,t)\ p(1 0,t)}
\end{align}
In order to determine $p(000,t+1)$ according to Eq.  (\ref{eq:chap2}), we have to sum up over the 32 different quintuplet configurations. 
And we have to follow this procedure 8 times, because there are $2^{3}=8$ possible triplet configurations. 
At the end, we arrive at a new  triplet distribution, which serves as the starting point for the next generation, and so forth. 

Regarding the computational effort, it is comparably low because of the straight forward calculation. 
Because in the VM  the  transition rates do not depend on the lattice position $i$ but rather on the local frequencies $f_{i}$,
it  would  be even possible to reduce the higher order approximation to a framework of frequencies of motives $\{000\}$, $\{001\}$, etc.
But to test the resulting improvement in computation speed is not the aim of this paper.

For the \emph{pair approximation}, we repeat again that we have to solve a system of three coupled equations  for $\mean{x}$, $c_{1|1}$ and $\mean{x_{1|1}}$.The Appendix provides explicit expressions for the involved quantities, where the transition rates are obtained from Table \ref{tab:trans}.  

\subsection{Prediction of the stationary solution}
\label{sec:stat}

As a first test of the validity of our approximations, we focus on the correct prediction of the stationary solution, $\mean{x^{\mathrm{stat}}}$. 
We have two parameters to vary: (i) the initial condition $x(t=0)$ which is changed from 0.1 to 0.9 in steps of 0.1, 
(ii) the parameter $\alpha$ that decides about majority and minority voting. 

\begin{figure}[htbp]
  \begin{center}
    \includegraphics[width=7.0cm,angle=0]{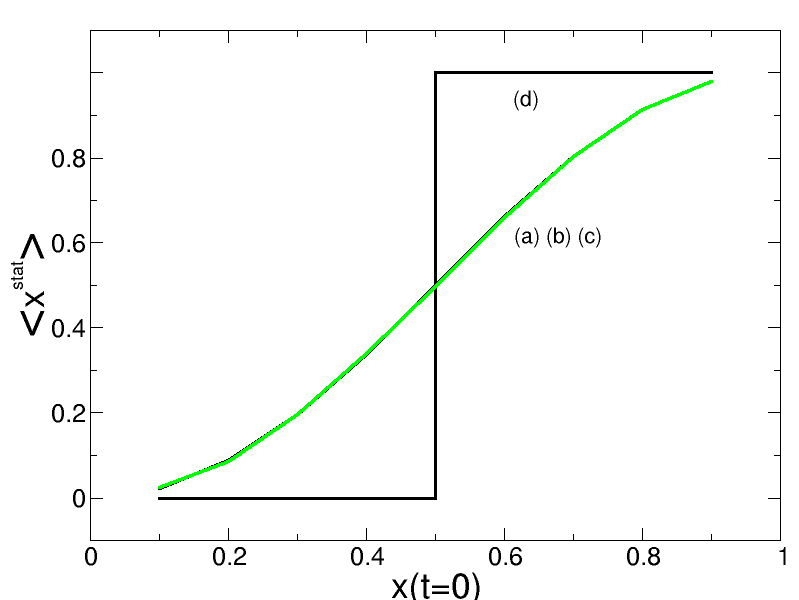} \qquad
    \includegraphics[width=7.0cm,angle=0]{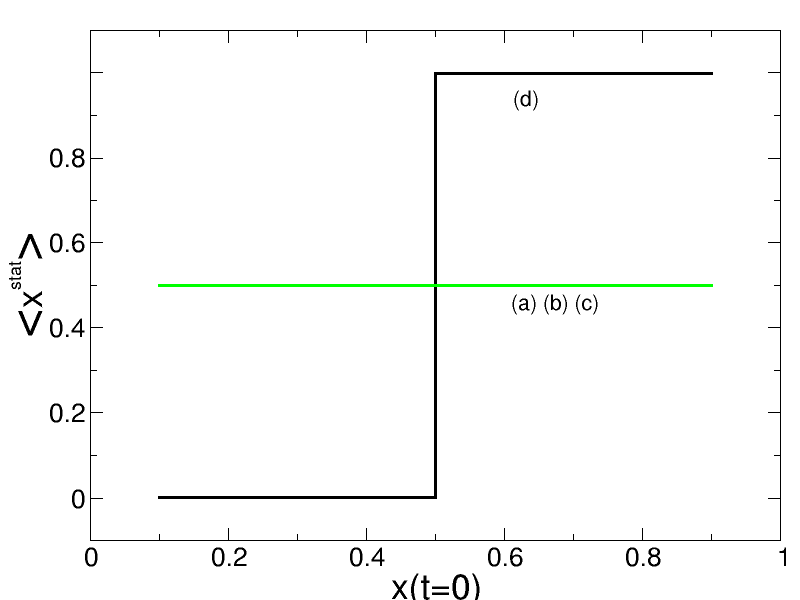}
    \caption{Stationary solution $\mean{x^{stat}}$ for majority voting. (left) deterministic case, $\alpha=0$, $\varepsilon=0$, (right) stochastic case, $\alpha=0.2$, $\varepsilon=10^{-4}$. Curves from different approaches: (a) CA simulation (reference case, green), (b) quintuplet approximation, 
      (c) pair approximation, (d) mean-field approximation}
    \label{fig:pos1dstat}
  \end{center}
\end{figure}

For the majority voting, the results are shown in Fig.~\ref{fig:pos1dstat} for the deterministic ($\alpha=0$) and the stochastic ($\alpha=0.2$) case. 
There are two remarkable observations. 
First, there is a noticeable difference between the deterministic and the stochastic outcome. 
In the latter, for the reference case, we always observe $\mean{x^{\mathrm{stat}}}=0.5$, regardless of the initial condition. 
This is understandable also from  Fig. \ref{fig:ca1dp} (lower left), where we observe the \emph{correlated coexistence} of the two opinions. 
The small disturbance $\varepsilon=10^{-4}$ acts as a repulsion from consensus and averaging over the large opinion domains in the long-term limit leads to the balance between the two opinions. 
In the deterministic case, however, the dynamics freezes really fast, within the first 20 generations, which is also shown in Fig. \ref{fig:ca1dp} (upper left).
Hence, not many deviations from $x(t=0)$ can happen. But $\mean{x^{\mathrm{stat}}}$ is slightly below (for $x(t=0)<0.5$) or above (for $x(t=0)>0.5$) this value, which indicates the trend towards consensus. 

The second remarkable observation is the difference between the mean-field approximation (d) and the other approximations (b), (c) in predicting the stationary value. 
Regardless of the details of the dynamics, the mean-field approximation always predicts \emph{consensus} for the opinion that was initially the majority, i.e. $x^{\mathrm{stat}}=0$ for $x(t=0)<0.5$ and vice versa. 
The quintuplet and the pair approximation, however, correctly predict the expected stationary solution, as the match with the reference case (a, green curve) indicates. 

Hence, we can conclude that the mean-field approximation is not suitable to describe the dynamics of the one-dimensional CA for majority voting, which is not very surprising. 
But we are more interested to find out to what extend the pair and the quintuplet approximation, which \emph{both} predict the expected \emph{stationary} solution correctly for both the deterministic and the stochastic case, are also able to describe the \emph{dynamics toward the stationary state} correctly. 
This is discussed in the following section. 

We note that the \emph{minority voting} rule, not further discussed here, always leads to  symmetric coexistence of both opinions, $x^{stat}=0.5$, for each initial condition. 
This was already indicated in Fig.~\ref{fig:ca1dp} (right column),. 
For minority voting, the mean-field approximation indeed predicts the correct stationary value, $\mean{x}=0.5$, but it miserably fails for majority voting.

\subsection{Prediction of  the time-dependent solution}
\label{sec:pred-time-depend}

\begin{figure}[htbp]
  \begin{center}
    \includegraphics[width=7.0cm,angle=0]{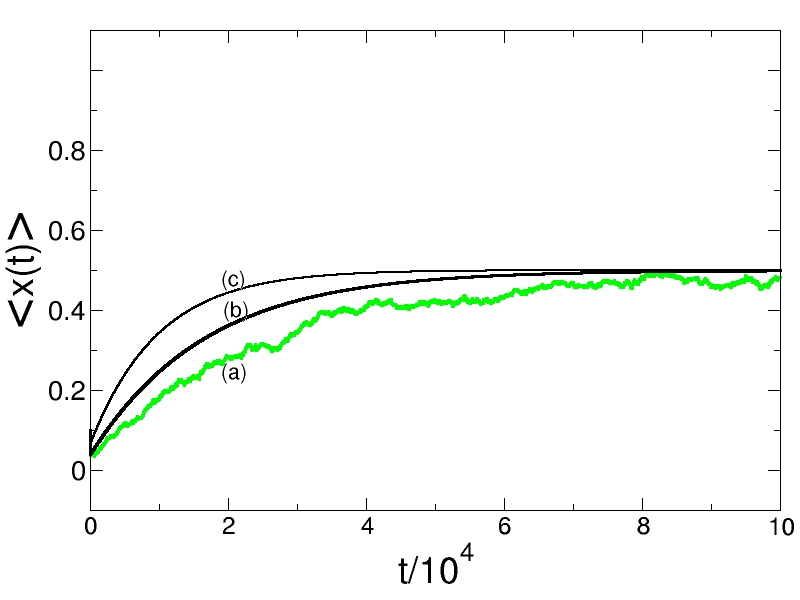} \qquad
    \includegraphics[width=7.0cm,angle=0]{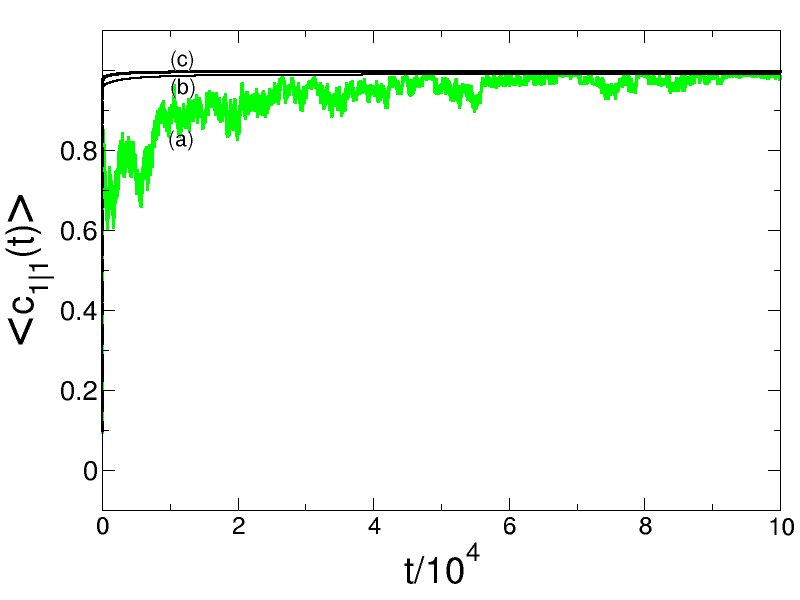}
\bigskip
  \includegraphics[width=7.0cm,angle=0]{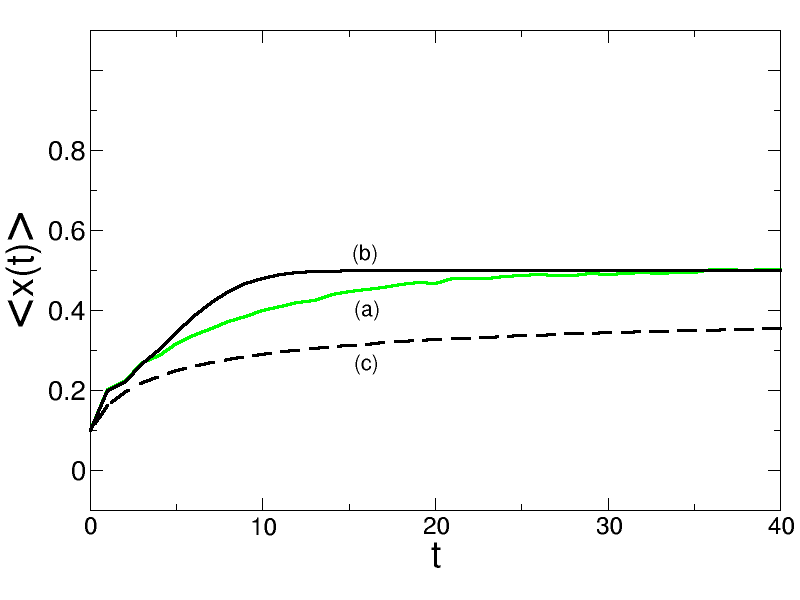}\qquad
   \includegraphics[width=7.0cm,angle=0]{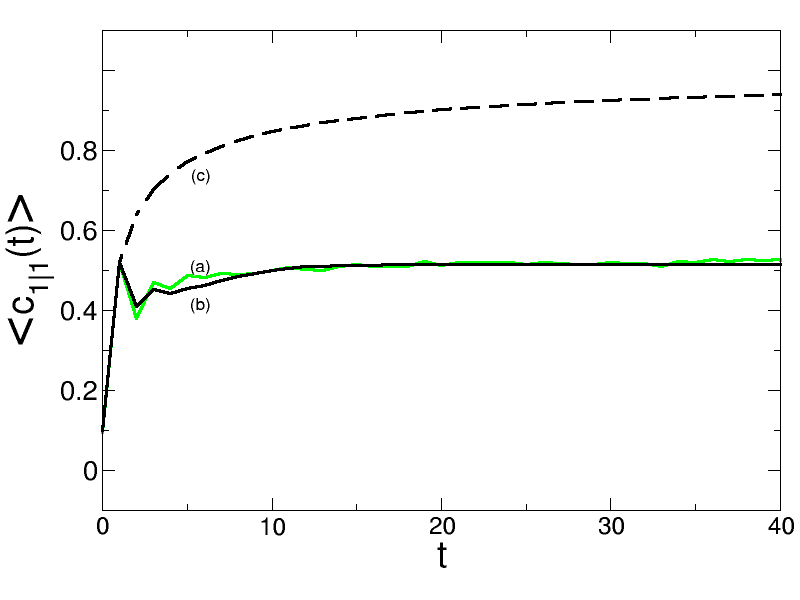}
    \caption{Dynamics of the expected frequency $\mean{x(t)}$ and the correlation $c_{1|1}(t)$. 
(upper row) majority voting, parameters:  $\alpha=0.2$, $\varepsilon=10^{-4}$, $x(t=0)=0.1$, $c_{1|1}(t=0)=0.1$. (lower row) minority voting, parameters: 
 $\alpha=0.8$, $\varepsilon=10^{-4}$, $x(t=0)=0.1$, $c_{1|1}(0)=0.1$. 
Curves from different approaches: (a) CA
      simulation (reference case, green); (b) quintuplet approximation; (c) pair approximation. }
  \label{fig:pos1df}
 \end{center}
\end{figure}

Again, we are interested in the dynamics of the \emph{expected} fraction of opinion 1, $\mean{x(t)}$.
We concentrate only on the stochastic dynamics, since the deterministic dynamics freezes very fast. 
In Fig. \ref{fig:pos1df}, we compare the validity of our two different approximations, the pair and the quintuplet approximation, with the reference case, the stochastic CA simulations, for both the majority and the minority voting rule. 

For the case of \emph{majority voting}, all the three curves for $\mean{x(t)}$ actually start at $x=0.1$, but then decrease rapidly during the first few generations (which cannot be noticed on the long time scale), before they steadily evolve toward the quasistationary state $\mean{x}=0.5$.
This kind of overshooting in the very initial phase can be also observed for other extreme initial conditions, e.g. $x=0.9$. 
In this case  $\mean{x(t)}$ will first increase before settling at $\mean{x}=0.5$. 
We see that the two approximations reach this state much faster than the simulations.  
While this gap  is noticable in particular in the beginning, the quintuplet approximation fares better than the pair approximation. 
The evolution of the
correlation term $c_{1|1}(t)$ starts at $c_{1|1}=0.1$ and reaches values near one very quickly. 
This indicates the formation of large domains of the same opinion, i.e. domains with different opinions are well separated, but can coexist.
However, we should notice that the stochastic simulations do not reach the value of 1, because there is never a real \emph{consensus}. 
Again, in the beginning, there is some discrepancy between the approximations and the simulations, but the quintuplet approximation is slightly more accurate. 

For the case of \emph{minority voting}, we notice that the quintuplet approximation fares considerably better than the pair approximation. 
The latter fails to predict the dynamics for both $\mean{x(t)}$ and $\mean{c_{1|1}(t)}$, whereas the quintuplet approximation is almost exact, except for a short time interval after the beginning, where it predicts a faster relaxation for   $\mean{x(t)}$. 
Values of $\mean{x}=0.5$ and $\mean{c_{1|1}}=0.5$ clearly indicate a random pattern as also shown in the snapshots of Fig. \ref{fig:ca1dp}.

\section{Conclusions}

With our investigations of the one-dimensional voter model, we want to achieve two goals: 
(i) a better understanding of the \emph{non-linearity} in the voter dynamics that was mostly studied as a linear model, only, 
(ii) a probabilistic description, and possible approximations, of the dynamics for the fraction of opinion 1, $x(t)$.

The non-linearity can be easily expressed by means of a free parameter $\alpha$.
The critical value $\alpha^{c}=1/3$ distinguishes between two different rules, \emph{majority} voting ($\alpha>\alpha^{c}$) and \emph{minority} voting ($\alpha<\alpha^{c}$), whereas $\alpha=\alpha^{c}$ refers to the border case of the \emph{linear} voter model. 
It is known from well-mixed populations that majority voting should result in \emph{consensus}, i.e. the asymptotic dominance of only one opinion, whereas minority voting should result in \emph{coexistence}, i.e. the occurrence of both opinions in different fractions. 

Our main focus was the role of local correlations in determining this outcome. 
For this we have used one-dimensional cellular automata (CA) in which each cell $i$, characterized by its opinion $\theta_{i}$, has a defined neighborhood of two cells with possibly different opinions, denoted as \emph{triplet}.
The transition probability of a cell to change its opinion is then determined by the local \emph{frequency of opinions} in its neighborhood (including the focal cell)  and the \emph{non-linear response} to this information, expressed by means of $\alpha$. 
The values of $\alpha$ define a certain probability to switch to the opposite opinion, i.e. we can use them to switch between a deterministic ($\alpha=0, 1$) or a stochastic  ($\alpha\neq 0, 1$) dynamics. 
In the latter case, we have further assumed a very small probability $\alpha_{0}\equiv \varepsilon= 10^{-4}$ to perturb a state of complete consensus, which allows a non-stationary dynamics of the CA as shown in Figure \ref{fig:ca1dp} (bottom right). 

While the minority rule only results in \emph{random coexistence} of the two different opinions, the majority rule generates more interesting results. 
In particular, we observe a \emph{correlated coexistence} characterized by the formation of large domains of the same opinion which  change continuously.
I.e., we have a non-equilibrium dynamics in which each opinion, for a certain time, can form  large clusters of the majority opinion. 

The question then is how to describe this dynamics mathematically.
In this paper, we follow a probabilistic approach, i.e. each cell has a certain probability of a given opinion $\sigma \in \{0,1\}$, which also depends on the probabilities of the nearest neighbors, second-nearest neighbors, and so forth. 
In order to close the dynamics, we have proposed three different approximations at different levels of the description.

The first level is the aggregated description in terms of the global fraction of opinion 1, $x(t)$, for which we derive a dynamics for the \emph{expected value}, $\mean{x(t)}$. 
On this level, we discuss two approximations. 
The simplest one is the \emph{mean-field approximation}, in which \emph{no correlations} between neighboring states are considered. 
So, we call this the zero-order approximation. 
It gives us a prediction for $\mean{x(t)}$ derived from the well-mixed case. 
In contrast, the \emph{pair approximation} considers a correlation between a cell and its neighbor, i.e. the triplet consisting of a cell and its two neighbors is decomposed in two cell-neighbor pairs. 
Correlations between neighbors are not considered, so we call this the 1st-order approximation. 
Hence, we have a prediction for $\mean{x(t)}$ coupled to the dynamics of the pair correlations $c_{1|1}(t)$. 

The third approximation does not refer to the aggregated level, but to the stochastic dynamics of a triplet, i.e. a cell with its two neighbors, that is determined by the larger neighborhood of a quintuplet, i.e. considers also the second nearest neighbors of the cell. 
Therefore, we call this the 2nd-order approximation.
Using certain assumptions, we are able to provide a \emph{closed form dynamics} for this larger neighborhood in terms of a probabilistic equation. 

To compare the validity of these mathematical approximations, we use as a \emph{reference case} stochastic computer simulations of the one-dimensional CA, which are averaged over a larger number of runs. 
We have discussed the majority and the minority voting, as well as the deterministic and the stochastic dynamics, separately. 

In conclusion, we can summarize that the zero-order approximation only predicts the stationary outcome of the minority voting correctly, but fails for the majority voting rule, both with respect to the dynamics and the stationary outcome. 

The first-order approximation performs comparably well in comparison to the second-order approximation only for the majority voting.  
The stationary outcome is correctly predicted both for the deterministic and the stochastic case, also the dynamics is covered fairly good. 
However, the first-order approximation fails to predict the dynamics of the minority voting. 
This case is only well covered by the second-order approximation that gives not only a correct description of $\mean{x(t)}$, but also of the pair correlations $\mean{c_{1|1}(t)}$. 

Commenting specifically on the correlations, we recall again that \emph{both} the first and second-order approximations lead to comparably good asymptotic results only for the majority voting. 
But they clearly predict a faster formation of domains, i.e. a convergence to their stationary value, as compared to the CA simulations. 
This limits their usability to fully understand the emergence of long-range correlations. 
In the case of minority voting, our first-order approximation fails, while the second-order approximation for $\mean{c_{1|1}(t)}$ could be even seen as accurate in its computational prediction. 
This is not so surprising if we recall that minority voting rules, different from majority voting, do \emph{not} result in long-range correlations.

Hence, we can conclude that the  quintuplet approximation that covers also the second-order neighborhood, is accurate enough to describe the dynamics of the CA on the macroscopic level. 
While it is of course understandable, that an approximation that considers more information is usually more accurate, we should also relate this conclusion to the computational effort. 
Here, it turns out that the 2nd-order approximation, although stochastic, i.e. needs to be averaged over a number of runs, performs very fast because of the closed form dynamics. 
Of course, the 1st-order approximation, the closed form of which is given in the Appendix, is computationally even simpler.
But there is a trade-off with the accuracy of the prediction. 
Still, as long only majority voting is considered, the 1st-order approximation should be preferred, both for simplicity, accuracy and computational effort. 

Our last remark is about the \emph{coexistence} of the two opinions, which is the more interesting scenario compared to \emph{consensus}, i.e. the existence of only one opinion. 
Here, we are not so much interested in the trivial case of \emph{random coexistence} without any structure formation, which is characterized by $\mean{x}=0.5$ and $c_{1|1}=0.5$. 
We focus more on  the case of \emph{correlated coexistence} which has an interesting complexity because of the formation of local structures, nicely shown in Fig. \ref{fig:ca1dp} (lower left). 
On the level of our approximations, this state is characterized by $\mean{x}=0.5$ and $c_{1|1}\to 1$. 
I.e., both opinions form at times large domains, indicated by the high pair correlation, but none of the two opinions entirely dominates the dynamics, as the expected value for its fraction is about 0.5.

Such insights can be generalized to other cases of frequency dependent processes which e.g. play a role in population ecology (invasion or extinction of species). 
For most of these applications a two-dimensional CA is more appropriate as it was discussed in \citep{Schweitzer2009}.
The one-dimensional CA investigated here, on the other hand, allows a mathematical approximation of the stochastic dynamics in terms of the 2nd-order neighborhood, which gives a much higher predictive power. 
This formalism can be also used for other frequency dependent processes in one-dimensional CA.

\subsection*{Acknowledgments}

The authors acknowledge fruitful discussions with H. M\"uhlenbein on an early version of this paper.

\section*{Appendix}

In \citep{Schweitzer2009}, we describe a method to calculate $\mean{x}$, $c_{1|1}$ and $\mean{x_{1,1}}$ for a two-dimensional non-linear VM. 
Here, we apply this approach to the one-dimensional CA, i.e. we consider a cell and its two nearest neighbors in pair approximation, using
the transition rates of Eqn. (\ref{trans2}). For simplicity, we use again the 
notation $x\equiv \mean{ x}$. With  Eqn. (\ref{eq:cond_prob}), we find for Eqn. (\ref{macro-pair})
\begin{equation}
\label{eq:master1d}
\begin{aligned}
  \frac{dx}{dt}=\varepsilon\left[\frac{1}{1-x}(1-2x+xc_{1|1})^2-x{c_{1|1}}^2 \right]+\frac{(1-3\alpha)x(2x-1)(1-c_{1|1})^2}{1-x}
\end{aligned}
\end{equation}
We note that $c_{1|1}=x$ if all neighboring states are uncorrelated, in
this case Eqn. (\ref{eq:master1d}) reduces to the mean-field
Eqn. (\ref{eq:mean_field}).

In order to calculate the time derivative of the doublet frequency
$\mean{ x_{1,1} }$ we have to consider how it is affected by changes of
$\sigma$ in a specific opinion pattern $\{\sigma,\sigma_{1},\sigma_{2}\}$,
where the $\sigma_{j}$ refer to the given states of the $m=2$ neighbors of a cell. 
In a frequency dependent process it is assumed that the transition
does not depend on the exact distribution of the $\sigma_{j}$, but 
only on the frequency of a particular state $\sigma$ in the neighborhood.
Let $S_{\sigma,q}$ describe a
neighborhood where the center cell in state $\sigma$ is surrounded by
$q$ cells of the same state $\sigma$.
For any given $q\leq m$, there
are $m \choose q$ different opinion patterns with the same value of $q$. The
global frequency to find a neighborhood  $S_{\sigma,q}$ is denoted
as  $x_{\sigma,q}$ with the expectation value
$\mean{x_{\sigma,q}}$. 
Regarding the possible transitions, we are only interested in changes of
the doublet (1,1), i.e. transitions $(1,1)\to(0,1)$ or $(0,1)\to(1,1)$.
The transition rates shall be denoted as $w\big((0,1)|(1,1),S_{\sigma,q}\big)$ and
$w\big((1,1)|(0,1),S_{\sigma,q}\big)$ respectively, which of course depend on the
local neighborhood $S_{\sigma,q}$. The dynamics of the expected
doublet frequency can then be described by the rate equation:
\begin{equation}
\label{d_frequency2}
\begin{aligned}
\frac{d}{dt}\mean{x_{1,1}}(t)= \sum_{m=0}^{n-1} \Big [
w\big((1,1)|(0,1),S_{0,q}\big)\mean{x_{0,q}} -
w\big((0,1)|(1,1),S_{1,q}\big)\mean{x_{1,q}} \Big]
\end{aligned}
\end{equation}
In order to specify the transition rates of the doublets
$w\big((\sigma^{\prime},1)|(\sigma,1),S_{\sigma,q}\big)$, with
$\sigma^{\prime}=1-\sigma$ and $\sigma=\{(0,1)\}$, we note that there are
only 6 distinct configurations of the neighborhood. Let us take the
example $\underline{\sigma}^{0}=\{1,1,1\}$.  A transition of the center cell
$1\to 0$ would lead to the extinction of 2 doublets
$\sigma,\sigma_{j}=\{1,1\}$. On the other hand, the transition rate of
the center cell is $\varepsilon$ as known from Eqn. (\ref{trans2}).  This would
result in $w\big((0,1)|(1,1),S_{1,2}\big)\propto 2\varepsilon$. For a lattice
of size $N$ the number of doublets is $N$, so also the number of
neighborhoods $\underline{\sigma}^{0}$ is $N$. Therefore, if we apply the transition
rates of the single cells, Eqn. (\ref{trans2}) to the transition of the
doublets, their rates remain unchanged. Similarly, if we take the
example $\underline{\sigma}^{0}=\{0,1,1\}$, a transition of the center cell
$0\to1$ would occur at the rate $1-\alpha$ and would create 2 new
doublets. We verify that
$w\big((1,1)|(0,1),S_{0,0}\big)=2 \;(1-\alpha)$. This way we can also
determine the other possible transition rates:    
\begin{equation}
\label{eq:tp_1da}
\begin{aligned}
\omega(01|11,S_{1,2})&=2\varepsilon \quad &
\omega(01|11,S_{1,1})&=\alpha\\
\omega(01|11,S_{1,0})&=0 & 
\omega(11|01,S_{0,2})&=0\\
\omega(11|01,S_{0,1})&=\alpha &
\omega(11|01,S_{0,0})&=2(1-\alpha)
\end{aligned}
\end{equation}  
Note that two of the transition rates are zero, because the respective
doublets (1,1) or (0,1) do not exist in the assumed neighborhood.

Eventually, we express $\mean{x_{\sigma,q}}$ in Eqn. (\ref{d_frequency2}) by 
\begin{equation}
  \label{s-q}
 \mean{x_{\sigma,q}}= \sum_{\underline{\sigma}^{\prime}}
  \mean{x_{\sigma,\underline{\sigma^{\prime}}}}
\end{equation}
and apply the  pair approximation, Eqn. (\ref{eq:pairapprox}), to $\mean{x_{\sigma,\underline{\sigma}^{\prime}}}$,  
to obtain  the dynamic equation for $\mean{x_{1,1}}$:
\begin{equation}
\label{eq:doubletg}
\begin{aligned}
\frac{d\langle x_{1,1}\rangle}{dt}=&+ 2\alpha \frac{x}{(1-x)}(1-
c_{1|1})(1-2x+xc_{1|1})+ 2(1-\alpha)\frac{{x}^2}{
  (1-x)}(1-c_{1|1})^2\\&-2\varepsilon x {c_{1|1}}^2 -2\alpha x (1-c_{1|1}) c_{1|1}\end{aligned} 
\end{equation}
Finally, we obtain for the change of the correlation
$c_{1|1}$ using Equations \eqref{eq:local1}, \eqref{eq:master1d}, and
\eqref{eq:doubletg}:  
\begin{equation}
\label{eq:master1dc}
\begin{aligned}
\frac{dc_{1|1}}{dt} =& -\varepsilon \left[\frac{c_{1|1}}{x(1-x)}
  (1-2x+xc_{1|1})^2 
+ {c_{1|1}}^3  - 2{c_{1|1}}^2\right] \\
   & + \frac{ (1-c_{1|1})^2\bigl( -2\alpha-c_{1|1}+3\alpha
  c_{1|1}-2(1-3\alpha)(1-c_{1|1})x\bigr)}{1-x}
\end{aligned}
\end{equation}
Thus equations \eqref{eq:master1d} and \eqref{eq:master1dc} are the final
closed form for the dynamics of the one-dimensional CA. 

If we set $\alpha=0,\varepsilon=0$, then we get following simple
macroscopic description for \emph{deterministic} majority voting:
\begin{eqnarray}
\label{eq:master1dp}
\frac{dx}{dt}&=&\frac{x}{1-x}\ (2x-1)(1-c_{1|1})^2 \\
\frac{dc_{1|1}}{dt}&=&\frac{(1-c_{1|1})^2}{1-x}\left[2x-c_{1|1}(2x-1)\right] \nonumber
\end{eqnarray}
\vfill

\end{document}